\documentclass{aastex63}

\shorttitle{DES-CNN}

\usepackage[utf8]{inputenc}
\usepackage[english]{babel}
\usepackage{graphicx}
\graphicspath{{./}{figures/}}
\usepackage{blindtext}
\usepackage{hyperref}
\usepackage{float}
\usepackage{subfiles} 
\usepackage{scalerel}
\usepackage{tikz}
\usetikzlibrary{svg.path}

\newcommand{\orcid}[1]{\href{https://orcid.org/#1}{\textcolor[HTML]{A6CE39}{\aiOrcid}}}

\definecolor{orcidlogocol}{HTML}{A6CE39}
\tikzset{
    orcidlogo/.pic={
        \fill[orcidlogocol] svg{M256,128c0,70.7-57.3,128-128,128C57.3,256,0,198.7,0,128C0,57.3,57.3,0,128,0C198.7,0,256,57.3,256,128z};
        \fill[white] svg{M86.3,186.2H70.9V79.1h15.4v48.4V186.2z}
        svg{M108.9,79.1h41.6c39.6,0,57,28.3,57,53.6c0,27.5-21.5,53.6-56.8,53.6h-41.8V79.1z M124.3,172.4h24.5c34.9,0,42.9-26.5,42.9-39.7c0-21.5-13.7-39.7-43.7-39.7h-23.7V172.4z}
        svg{M88.7,56.8c0,5.5-4.5,10.1-10.1,10.1c-5.6,0-10.1-4.6-10.1-10.1c0-5.6,4.5-10.1,10.1-10.1C84.2,46.7,88.7,51.3,88.7,56.8z};
    }
}

\newcommand\orcidicon[1]{\href{https://orcid.org/#1}{\mbox{\scalerel*{
                \begin{tikzpicture}[yscale=-1,transform shape]
                \pic{orcidlogo};
                \end{tikzpicture}
            }{|}}}}

\begin{document}

\title{Identifying Transient candidates in the Dark Energy Survey using Convolutional Neural Networks}

\correspondingauthor{Venkitesh Ayyar}
\email{vayyar@bu.edu}

\author{\orcidicon{0000-0001-9081-1840} Venkitesh Ayyar}
\affiliation{Lawrence Berkeley National Laboratory, 1 Cyclotron Rd, Berkeley, CA, 94720, USA}
\affiliation{Hariri Institute for Computing and Computational Science and Engineering, Boston University, Boston, MA, 02215, USA}

\author{\orcidicon{0000-0002-3803-1641} Robert Knop Jr.}
\affiliation{Lawrence Berkeley National Laboratory, 1 Cyclotron Rd, Berkeley, CA, 94720, USA}

\author{\orcidicon{0000-0001-5287-3004} Autumn Awbrey}
\affiliation{Lawrence Berkeley National Laboratory, 1 Cyclotron Rd, Berkeley, CA, 94720, USA}
\affiliation{Department of Astronomy, University of California, Berkeley, Berkeley, CA, 94720, USA}

\author{\orcidicon{0000-0001-8983-4893} Alexis Andersen} 
\affiliation{Lawrence Berkeley National Laboratory, 1 Cyclotron Rd, Berkeley, CA, 94720, USA}
\affiliation{Department of Astronomy, University of California, Berkeley, Berkeley, CA, 94720, USA}

\author{\orcidicon{0000-0002-3389-0586} Peter Nugent}
\affiliation{Lawrence Berkeley National Laboratory, 1 Cyclotron Rd, Berkeley, CA, 94720, USA}
\affiliation{Department of Astronomy, University of California, Berkeley, Berkeley, CA, 94720, USA}

\begin{abstract}
The ability to discover new transient candidates via image differencing without direct human intervention is an important task in observational astronomy. For these kind of image classification problems, machine Learning techniques such as Convolutional Neural Networks (CNNs) have shown remarkable success. In this work, we present the results of an automated transient candidate identification on images with CNNs for an extant dataset from the Dark Energy Survey Supernova program (DES-SN), whose main focus was on using Type Ia supernovae for cosmology. By performing an architecture search of CNNs, we identify networks that efficiently select non-artifacts (e.g. supernovae, variable stars, AGN, etc.) from artifacts (image defects, mis-subtractions, etc.), achieving the efficiency of previous work performed with random Forests, without the need to expend any effort in feature identification. The CNNs also help us identify a subset of mislabeled images. Performing a relabeling of the images in this subset, the resulting classification with CNNs is significantly better than previous results, lowering the false positive rate by 27\%  at a fixed missed detection rate of 0.05.
\end{abstract}


\section{Introduction} \label{sec:intro}

A major aspect of observational astronomy is the "survey" which involves the wholesale mapping of various regions of the sky to create catalogs which are subsequently mined for scientifically important astronomical objects. 
We refer to a {\it transient candidate} as the detection on a single image of a new or varying source with respect to a previously taken reference image, regardless of its astrophysical nature since at this stage its classification is unknown and will remain so until further data is taken (spectroscopy and/or additional photometry). Some examples of such transient candidates are solar system objects, supernovae, active galactic nuclei, variable stars, and neutron star mergers, etc. Since some of these events are quite rare and will fade rapidly, it is often important to trigger follow-up observations immediately to glean their underlying nature and discover new physics. Hence, identifying transient candidates in images quickly and efficiently is very important so as not to waste precious, and expensive, follow-up resources.
For many years this process was conducted by manual inspection of images by humans. However, given the magnitude of image data generated by modern telescopes, it became imperative to automate this process via machine learning techniques. This was first done by the SNFactory \citep{Bailey:2007qn} where boosted decision trees were employed to greatly reduce the number of candidates. Subsequently,  \cite{Bloom:2011as}  advanced this work using random forests to  explore the Palomar Transient Factory data for new transients. Surveys such as the Dark Energy Survey (DES) ~\citep{2005IJMPA..20.3121F} map the sky both on a large scale and deeply, producing up to 170GB of raw imaging data every night.

In this work, we describe our efforts to perform transient candidate identification in DES. This work builds on previous work with random forest~\citep{Goldstein_2015,Mahabal_2019,10.1093/mnras/stv292} to classify Type Ia supernova from other artifacts of processing and instrumentation for the DES-SN data. Machine learning techniques such as Convolutional Neural networks (CNNs)~\citep{726791} have shown remarkable success in image classification problems. Here we apply these to identify transient supernova images. 

CNNs have been used for transient candidate identification by multiple groups. The work of \cite{Cabrera} and \cite{Cabrera_deepHITS} were focused on $u$-band imaging from DECam. In \cite{gieseke}, they used data from the SkyMapper Survey and their true-positives were solely drawn from discovered supernovae. Several groups have undertaken the challenge of transient candidate discovery in very wide-field, under-sampled imaging for surveys like TESS, GOTO and GWAC \citep{Jayaraman_TESS,Killestein_GOTO,turpin_GWAC}. 

In \cite{gomez_TAOnet} they use both the spatio and {\it temporal} data (from a higher cadence survey) to train their CNN's. This extra data is invaluable for classifying many transient candidates, but lacks the flexibility to work with a single image. We are also aware of another unpublished work with CNNs with a DES dataset
\footnote{Previous work using CNN's on DES data can be found {\href{https://medium.com/dessa-news/space-2-vec-fd900f5566}{here}}}.
In \cite{Acero} they train CNNs for transient detection while focusing on avoiding the use of a difference image in the training process. In \cite{duev_ZTF}, the authors include galactic transients while training their CNNs.

Here, we use the dataset found in \cite{Goldstein_2015} where stamps were placed on galaxies, drawn from an appropriate redshift range for the survey, and within the galaxy at a location proportional to its surface brightness.  This is different than the approach in \cite{Cabrera} and \cite{Cabrera_deepHITS}, where the true-positives in their training and validation data were generated by selecting stamps of real PSF-like sources and placing them at a different location at the same epoch and in the same CCD they were
observed. This approach works well for $u$-band transients, as potential backgrounds are quite faint (i.e. a host galaxy). However, for a number of transient candidates in an optical search (AGN, supernovae, etc.) this is troublesome as their brightness is comparable to or fainter than their associated galaxies.

After giving a brief description of CNNs in Section \ref{sec:cnn}, we describe our dataset in Section \ref{sec:dataset}. In Section \ref{sec:results}, we discuss our procedure and present our results. Finally, we summarize our findings in Section \ref{sec:conclusions}. Our code is available at the \href{https://github.com/vmos1/cnn_supernova_b_paper}{github repository} \footnote{We provide the best saved models and notebooks for plotting and visualization \href{https://github.com/vmos1/cnn_supernova_b_paper/tree/master/model_inference}{here}}.


\section{Convolutional Neural Networks} \label{sec:cnn}

Here we give a brief introduction to convolutional neural networks. 
{\it Neural networks} are machine learning computing systems that are very efficient at learning patterns in input data.
These are generic functions consisting of weight parameters organized in layers. Acting on the input data, after periodic application of non-linear {\it activation} functions, they produce outputs which can be either numbers (for regression) or class labels (for classification). By minimizing the deviation between computed output and the expected output, one arrives at the optimal weight parameters.
The procedure to compute the weights of a network is called {\it training the network}. A properly trained network learns the generic function and can correctly predict the outputs for an unseen dataset. Essentially, they are universal function approximators. 

\begin{figure}[bh]
\centering
\includegraphics[width=16cm]{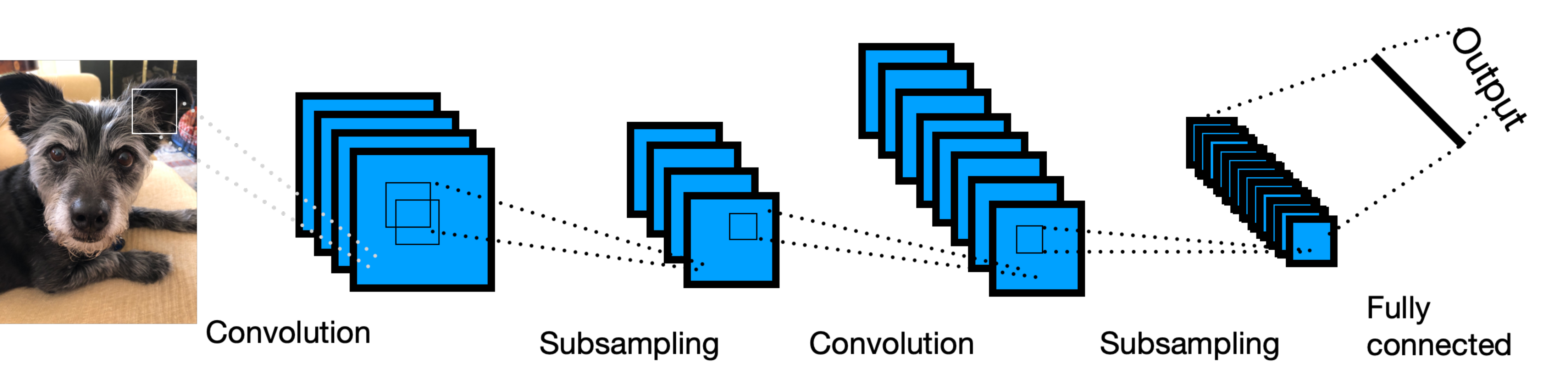}
\label{fig:cnn}
\caption{The general structure of a CNN with {\it convolution}, {\it subsampling} and {\it fully connected} layers.}
\end{figure}

Convolutional neural networks are a class of neural networks that specialize in recognizing patterns in image data. Using blocks of kernels that scan through the images, they extract features at different scales. Figure~\ref{fig:cnn} gives the general layout of a CNN. A typical CNN is made up of the following basic layers:

\begin{itemize}
    \item {\bf Convolution layers}:
    These perform convolutional operations on the images to extract feature maps.
    \item {\bf Subsampling layers}:
    Operating on feature maps, the subsampling layers compress the dimensionality to reduce the number of parameters. 
    \item {\bf Fully connected layers}:
    These layers combine different features of a single layer together.
\end{itemize}

CNNs typically have very large number of parameters and hence are prone to overfitting. One way to mitigate this is by using dropout layers that help suppress the unimportant weights by setting weight parameters to zero during training.

CNNs have been used extensively for image recognition, classification for images obtained both in the real world and in scientific experiments~\citep{726791,6248110}. Large multi-layered Neural networks, despite having large number of free parameters have shown remarkable success in image classification~\citep{GoogLeNet,alexnet,Bhimji:2017qvb}. Nevertheless, many studies have used specialized CNNs that have connections between non-adjacent layers such as {\it Resnet}~\citep{DBLP:journals/corr/HeZRS15} and {\it Unet}~\citep{DBLP:journals/corr/RonnebergerFB15}.
In a previous work~\citep{Ayyar:2020ijy}, it was found that instead of using CNNs with a few but long layers, CNNs with a resonably high number of layers with fewer parameters were remarkably successful in classifying signal from background for datasets in high energy physics experiments. This prompted us to explore the potential of such layered {\it deep} CNNs for classification problems in astronomy.

\section{Dataset} \label{sec:dataset}

\subsection{Dataset} \label{dataset}

We used the same dataset used in ~\cite{Goldstein_2015}.  The data collected is from the DES science operations from August 2013 to February 2014 and consists of 898,963 independent samples. Each sample in turn consists of 3 types of 2D images of dimension $51 \times 51$. The 3 types are labelled: {\it Template}, {\it Search}, and {\it Difference}. To incorporate the information from all 3 images, we use them as channels. In other words, each input sample is a 3 channel image of dimensions $51 \times 51$, having an expected label: Artifact (1) or Non-Artifact (0).
Due to the original timing of the data collection, it lacked non-artifact sources. Hence, the original authors used the method of artificial source construction, and thus injected these non-artifacts into the images. This method has been used extensively before ~\citep{Bailey:2007qn,Bloom:2011as}. For this dataset, all non-artifacts were artificially generated. 

\begin{figure}[ht!]
\plotone{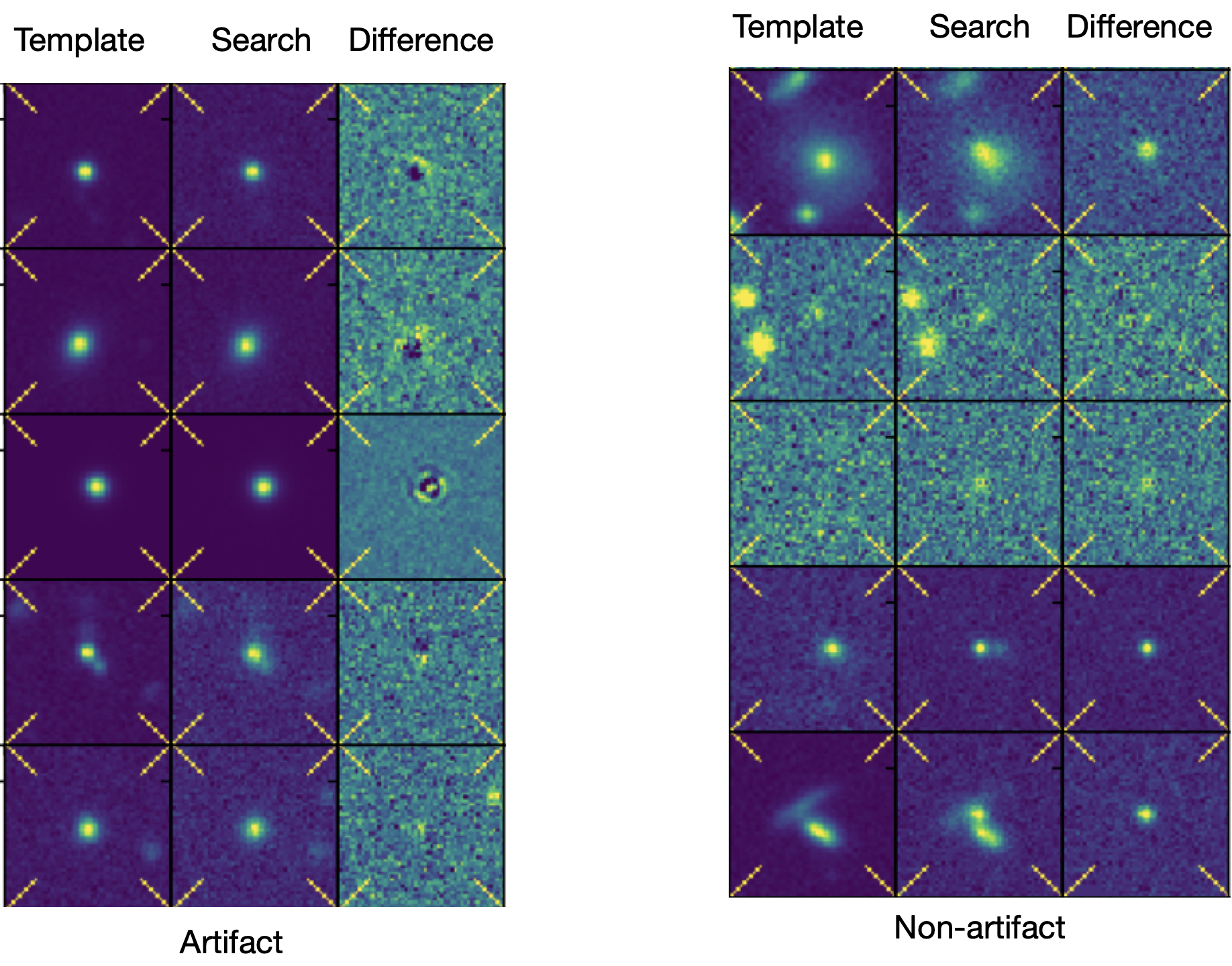}
\caption{The figure shows the three channels: template, search and difference for 5 different artifact and non-artifact samples.\label{fig:input}}
\end{figure}

Figure~\ref{fig:input} depicts the three channels for 5 independent samples for both artifacts and non-artifacts. Distinguishing them visually requires some level of expertise. A more detailed explanation of the dataset can be found in ~\cite{Goldstein_2015}.

\subsection{Classification and ROC curves}
\begin{figure}[ht!]
\epsscale{0.6}
\plotone{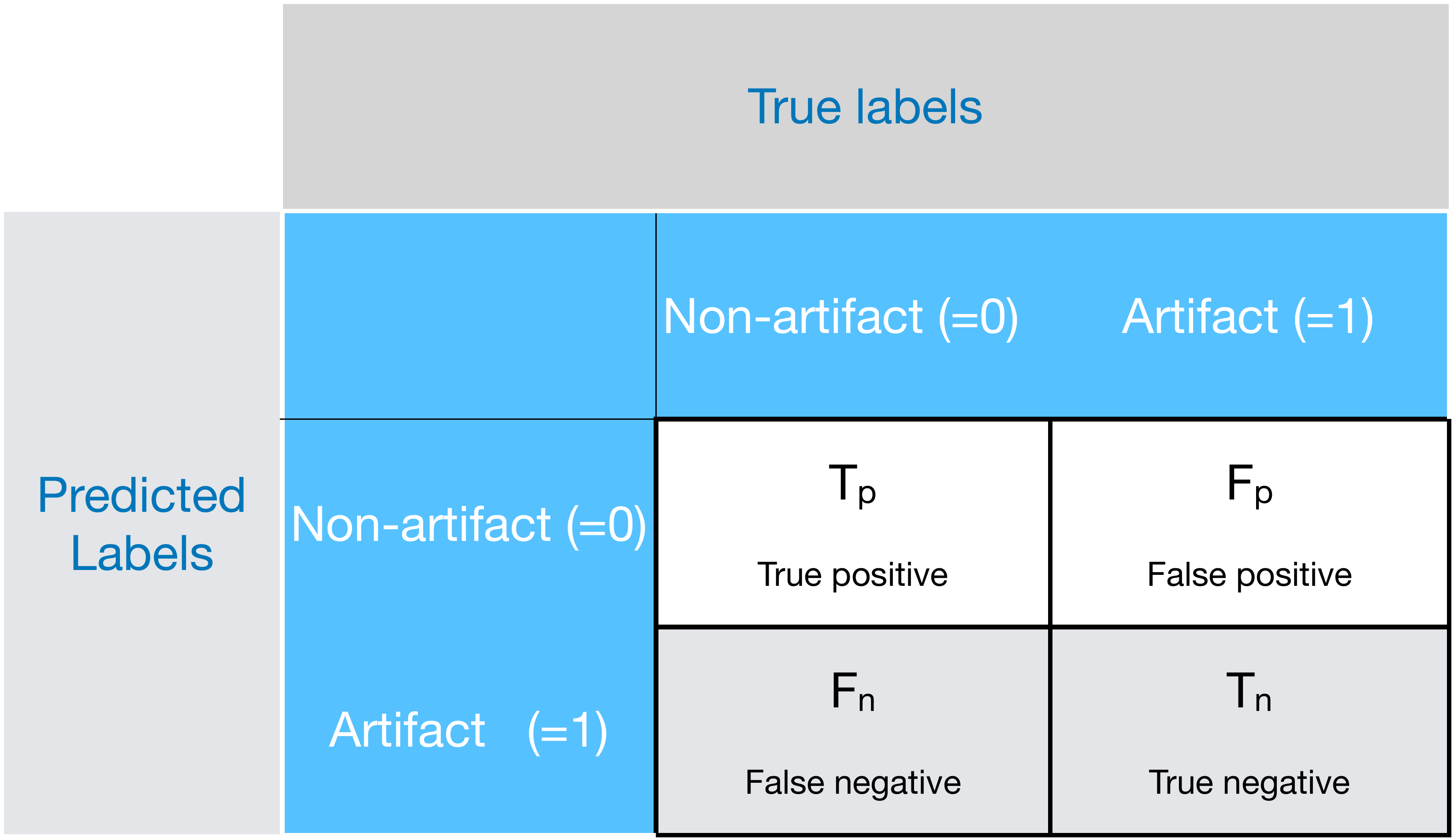}
\caption{The Confusion matrix for a binary classifier.\label{fig:cnf}}
\end{figure}

In classification problems, the model provides a class prediction for each sample. The aim is to develop a model that categorizes most samples correctly. For a binary classification problem like this one, the performance can be summarized by a  $2 \times 2 $ confusion matrix shown in Figure~\ref{fig:cnf}.
Some common classifier performance metrics are the True Positive Rate (TPR), False Positive Rate (FPR) and Missed Detection Rate (MDR). They are defined as: 
\begin{eqnarray}
    MDR&=&\frac{F_n}{T_p+F_n} \nonumber \\
    FPR&=&\frac{F_p}{F_p+T_n} \nonumber \\
    TPR&=&\frac{T_p}{T_p+F_n}  \label{eq:tpr}
\end{eqnarray}
where the quantities $ T_p $ , $ T_n $, $ F_p$ and $F_n$ are defined in Figure~\ref{fig:cnf}.
Since the classifier prediction is a floating point  number between 0 and 1, one uses a {\it threshold} parameter to determine a predicted class. The behavior of the classifier as the  threshold is varied, can be seen through the Receiving Operator Characteristic (ROC) curve. The ROC curve is the most commonly used method to compare a set of classifier models. A useful quantity used for comparison of models is the Area Under the Curve (AUC), which is expected to approach 1.

\section{Analysis and Results} \label{sec:results}

\subsection{CNN architecture search}

The goal of this work was to develop optimized CNNs by exploring various CNN architectures. 
Starting with 4-5 different CNN architectures, we obtained new architectures by varying kernel sizes, dropout layer locations, dropout ratios, types of pooling, learning rates, etc., and compared the classification performance of these models. Picking the best performing models among these, we performed further variations and compared their performance. After a few such iterations and studying about 100 different CNN models in total, we shortlisted 4 models with fairly different architectures that achieved good performance. The structures of these models are listed in Table~\ref{tab:model_structure}.

Some of the factors guiding the initial architectures were : 
\begin{itemize}
    \item Presence of Pooling layers (models 1-3 in Tables~\ref{tab:model_structure}) as opposed to the use of kernel striding (model 4) to reduce image size during convolution.
    \item Addition of multiple convolutional and batch normalization layer blocks before implementing pooling layers (models 2,3).
    \item Presence of dropout layers after batch normalization layers.
\end{itemize}

We explored kernel sizes from 2 to 8 (since the image size is 51 x 51) and convolutional layer sizes from 10 to 400. 
As we approached models with good classification performance, we lowered the learning rate to obtain better classification.
The entire code was written in python using the keras package~\citep{chollet2015keras}. We used the numpy library~\citep{harris2020array} for computations and jupyter notebooks~\citep{Kluyver2016jupyter} for analysis and visualizations.
While we do not claim to have explored every architecture, our search is reasonably thorough and we do find multiple CNN models that are efficient for the given classification problem.

\subsection{Results with original labels} \label{subsec:orig}
For the first part of the work, we split the data into {\it Training} (50\%), {\it validation} (5\%) and {\it test} samples (5\%). The validation data was used to assess the classification performance of trained models on unseen data. The test data was used to compare the ROC curves of the different models. At this stage, we kept about 40\% of the data in reserve for further analysis. We also trained a random forest using the hyperparameters described in the ~\cite{Goldstein_2015}. 

\begin{figure}[ht!]
\plotone{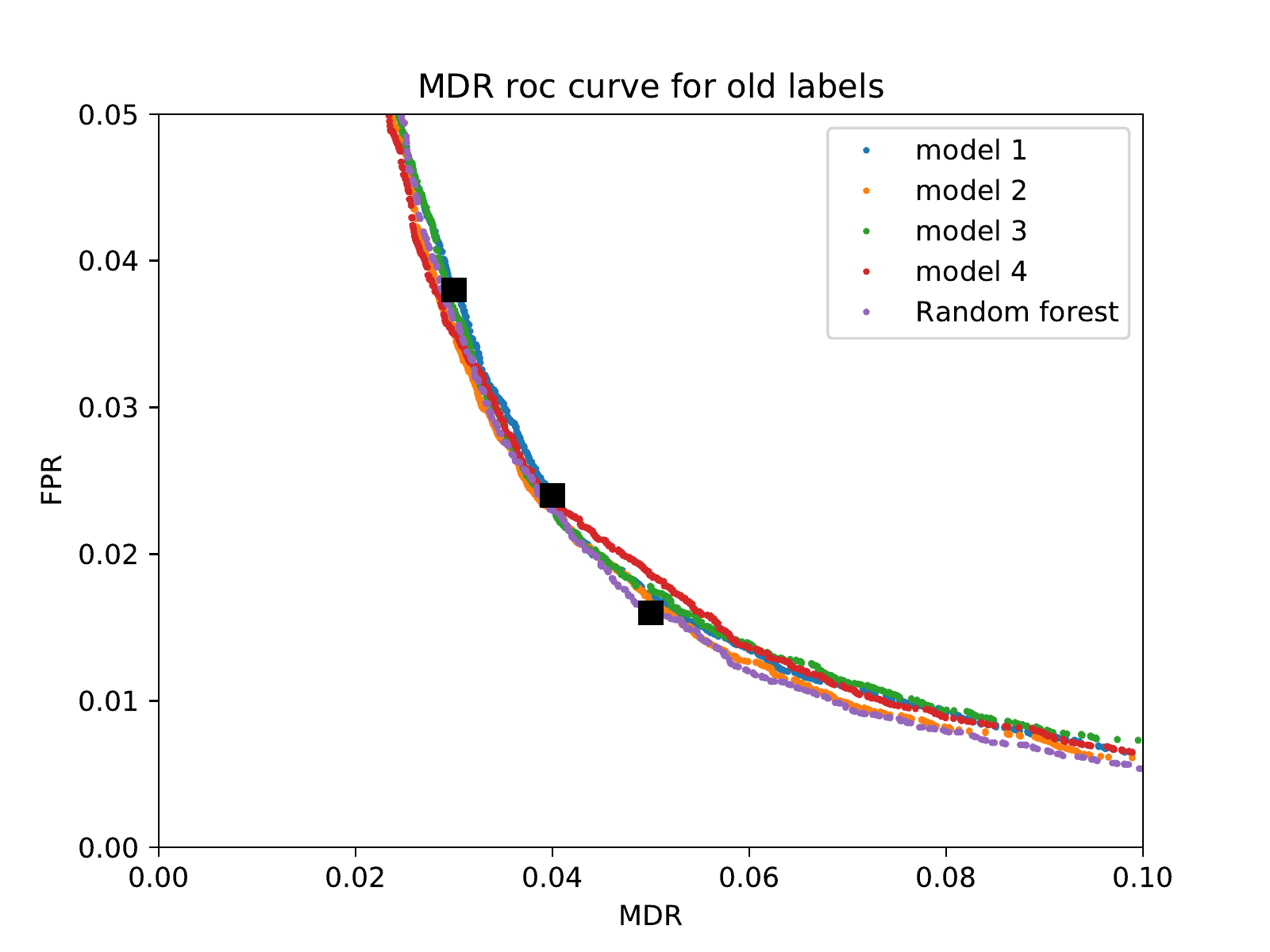}
\caption{The ROC curve of FPR vs MDR for the CNN models. The black squares show the points obtained in~\cite{Goldstein_2015} with a random forest. The ROC curves of all the models are adjacent to each other implying similar classification performance.}
\label{fig:old_roc}
\end{figure}

Table~\ref{tab:model_compare} compares the four best CNN models. All models have an AUC score close to 1.0.
Figure~\ref{fig:old_roc} shows the FPR-MDR ROC curve for the best CNN models. The black squares represent the ROC curve of the random forest from Figure~7 of ~\cite{Goldstein_2015}. Here we see that our CNN models and the random forest are comparable in performance to the previous work.

\begin{table}[h!]
\centering
\begin{tabular}{||c|c|c|c||}
\hline
Model name & Number of parameters & Area under curve (AUC) & Training time per epoch on GPU  \\
\hline
\hline
1 & 266k & 0.994 & 60s \\
2 & 415k & 0.994 & 127s \\
3 & 853k & 0.993 & 190s \\
4 & 954k & 0.994 & 47s \\
\hline
\end{tabular}
\caption{Table describing the best performing CNN models.}
\label{tab:model_compare}

\end{table}

\begin{figure}[ht!]
\plotone{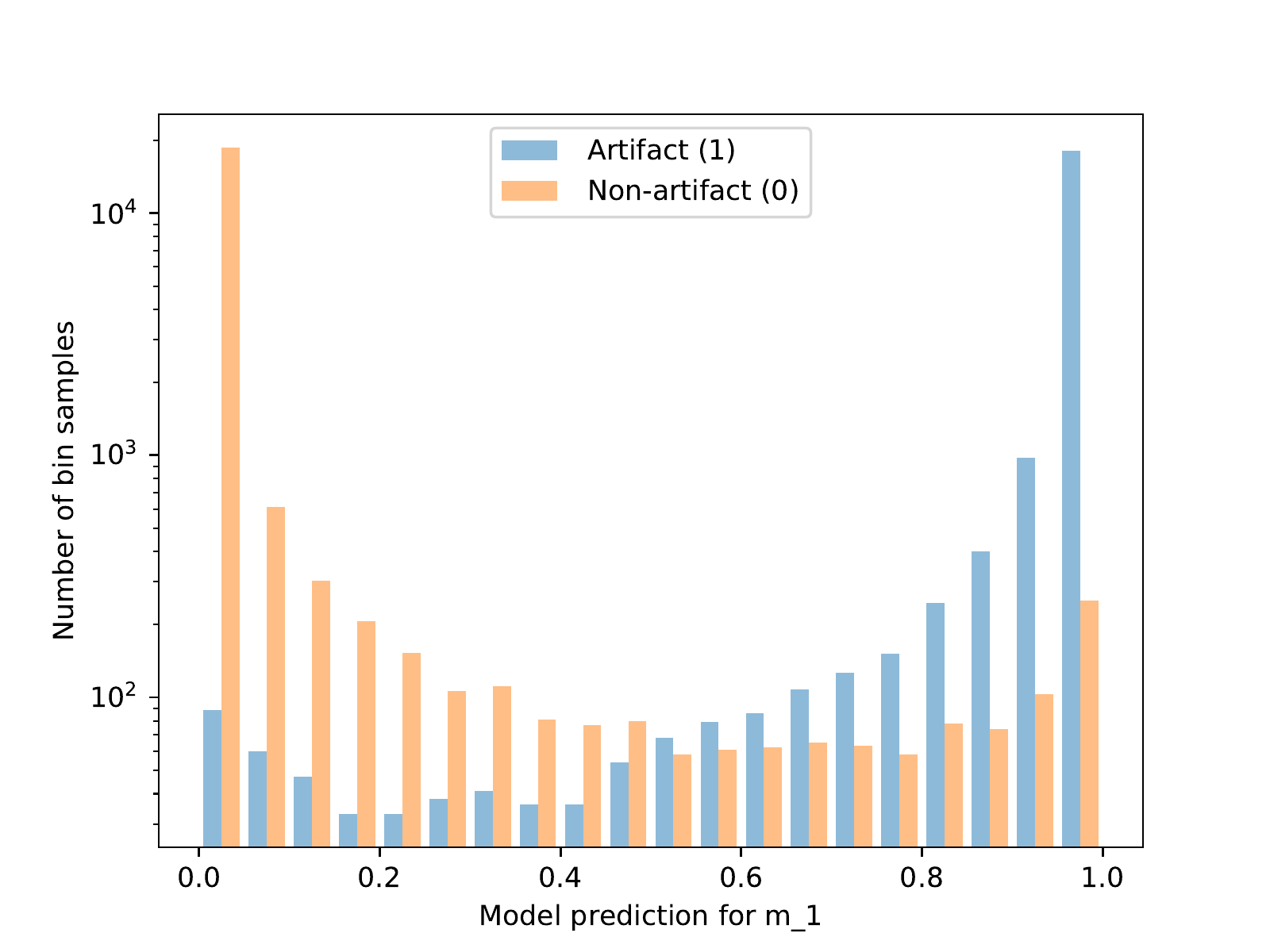}
\caption{The prediction histograms for Model 1.  Ideally, the classifier should have a prediction value 0 for all artifacts and 1 for all non-artifacts. Note the log scale on the y-axis. Most of the labels are predicted correctly. Also, there are relatively very few predictions in the intermediate region.}
\label{fig:hist1}
\end{figure}

Figure~\ref{fig:hist1} shows the prediction histogram for Model 1. Noting the log-scale on the y-axis, it is clear that most of the samples are classified correctly as either artifacts or non-artifacts. Also, very few samples have prediction scores in the intermediate 0.2-0.8 range, which is the hallmark of a good classifier. The prediction histograms for the other three models also look very similar.

\subsection{Mislabeled images}
Since the CNNs use the entire information from the images, one would expect well trained CNNs to achieve optimal classification performance. While the CNNs in Figure~\ref{fig:old_roc} do perform very well, the fact that they do not improve upon the performance of the random forest, prompted us to the explore their classification in more detail.

In the bottom left part of Figure~\ref{fig:hist1}, it can be seen that a significant number of artifacts with input label 1 are accorded a prediction value very close to 0. Similarly, a large number of non-artifacts with input label 0 have prediction values close to 1, as seen in the bottom right. We see a similar pattern for the other CNN models. This seems to imply that the CNN models are strongly mis-classifying a few samples, thus affecting the quality of their ROC curves.

To better understand this issue of strong mis-classification, we divided the samples into 6 categories depending on the original label and the predicted value for the model. For example, category 1 corresponds to samples that are labeled as artifacts (label=1), but have prediction values between 0 and 0.1.  We then compare the number of samples in the different categories for the different models. The results and description of the categories are summarized in Table~\ref{tab:categories}. It can be see than the CNN models have more points in categories 1 and 6, corresponding to strongly mis-classified samples.

\begin{table}[h!]
\centering
\begin{tabular}{|l||c|c|l|l|l|l|}
\hline
Category & Original Label & Prediction range & Description & Model 1 & Model 3 & Random Forest  \\
\hline
1 & 1 & 0-0.1 & Strongly mis-classified Artifact & 0.35 \% & 0.76 \% & 0.15 \% \\ 
2 & 1 & 0.1-0.5 & Weakly mis-classified Artifact & 0.9 \% & 1.2 \% & 1.0 \% \\ 
3 & 1 & 0.5-1.0 & Correctly classified  Artifact & 48.2 \% & 47.7 \% & 48.4 \% \\ 
\hline
4 & 0 & 0 - 0.5 & Correctly classified  Non-artifact & 48.4 \% & 49 \% & 48.4 \% \\ 
5 & 0 & 0.5 - 0.9 & Weakly mis-classified  Non-artifact & 1.2 \% & 0.8 \% & 1.6 \% \\ 
6 & 0 & 0.9 - 1.0 & Strongly mis-classified  Non-artifact & 0.8 \% & 0.7 \% & 0.4 \% \\ 
\hline
\end{tabular}
\caption{Dividing the samples into categories, depending on model predictions. Categories 1 and 6 correspond to the strongly mis-classified samples. The CNN models have more strongly mis-classified samples.}
\label{tab:categories}

\end{table}
To better understand this, we look at how the samples classified by Model 1 into category 1 are categorized by other models. In Table~\ref{tab:misclassified_comparison}, we show the prediction values for the other 3 models and random forest, for that were classified by Model 1 to be in categories 1 and 6. It can be seen that about 66\% of the samples are also placed in the same category by the other 3 CNN models. However, the random forest model places a much smaller proportion of these samples in category 1. This is further confirmation that the four CNN models seem to strongly mis-classify the same set of images.

\begin{table}[h!]
\centering
\begin{minipage}{.45\linewidth}
{\bf Points in category 1 for model 1} \\
\begin{tabular}{|l||c|c|c|}
\hline
Category & 1 & 2 & 3   \\
\hline
CNN 2 &  65.8 \% & 20.1 \% & 14.1 \% \\ 
CNN 3 &  72.5 \% & 18.8 \% & 8.7 \% \\ 
CNN 4 & 72.5 \% & 17.4 \% & 10.1 \% \\ 
Random forest & 26.8 \%  & 45.0 \% &  28.2 \% \\ 
\hline
\end{tabular}
\end{minipage}
\begin{minipage}{.5\linewidth}
{\bf Points in category 6 for model 1} \\
\begin{tabular}{|l||c|c|c|}
\hline
Category & 4 & 5 & 6   \\
\hline
2 &  7.9 \% & 18.4 \% & 73.7 \% \\ 
3 &  12.4 \% & 11.9 \% & 75.7 \% \\ 
4 & 9.9 \% & 29.3 \% & 60.7 \% \\ 
Random forest & 15.8 \%  & 44.6 \% &  39.5 \% \\ 
\hline
\end{tabular}
\end{minipage}
\caption{
The table on the left shows how the different models categorize the 149 samples of a test dataset that are in category 1 (artifact classified as non-artifact) for Model 1. The CNNs place over 66\% of these samples in category 1. The random forest places only about 27\% of these, instead placing many of the rest in category 2.
Similarly, the table on the right shows how the different models categorize the 354 samples of a test dataset that are in category 6 for Model 1.
From these, it can be inferred that, the different CNN models all strongly misclassify the same set of points.}
\label{tab:misclassified_comparison}
\end{table}

\subsection{Relabeling}
\begin{figure}[ht!]
\begin{minipage}{0.5\textwidth}
\includegraphics[width=.95\textwidth]{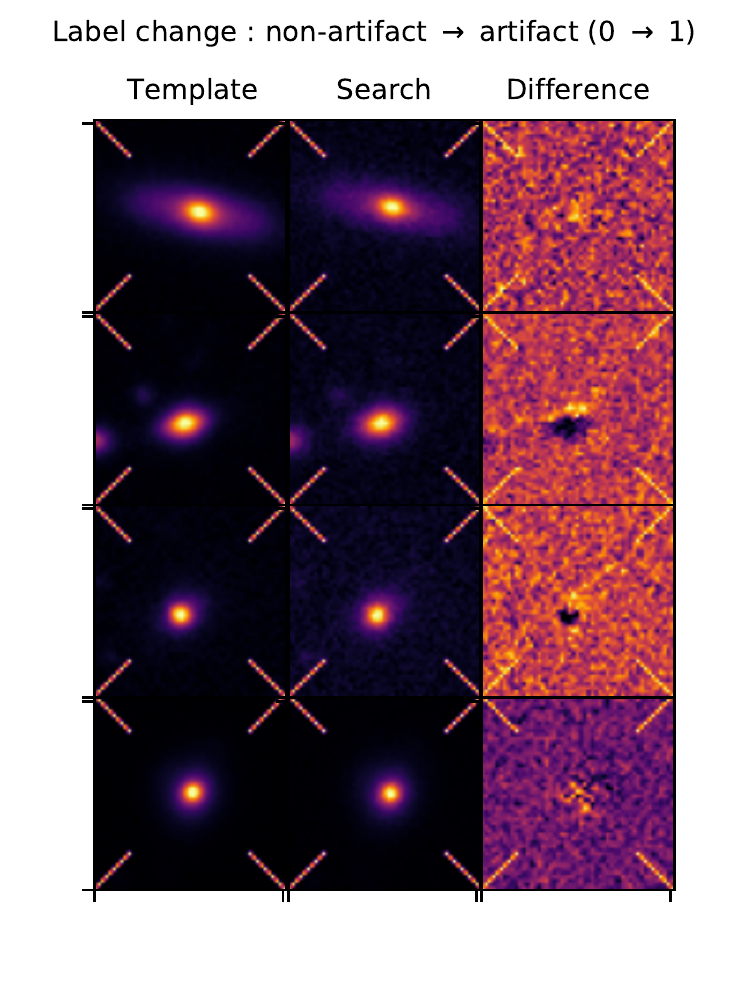}
\end{minipage}
\begin{minipage}{0.5\textwidth}
\includegraphics[width=.95\textwidth]{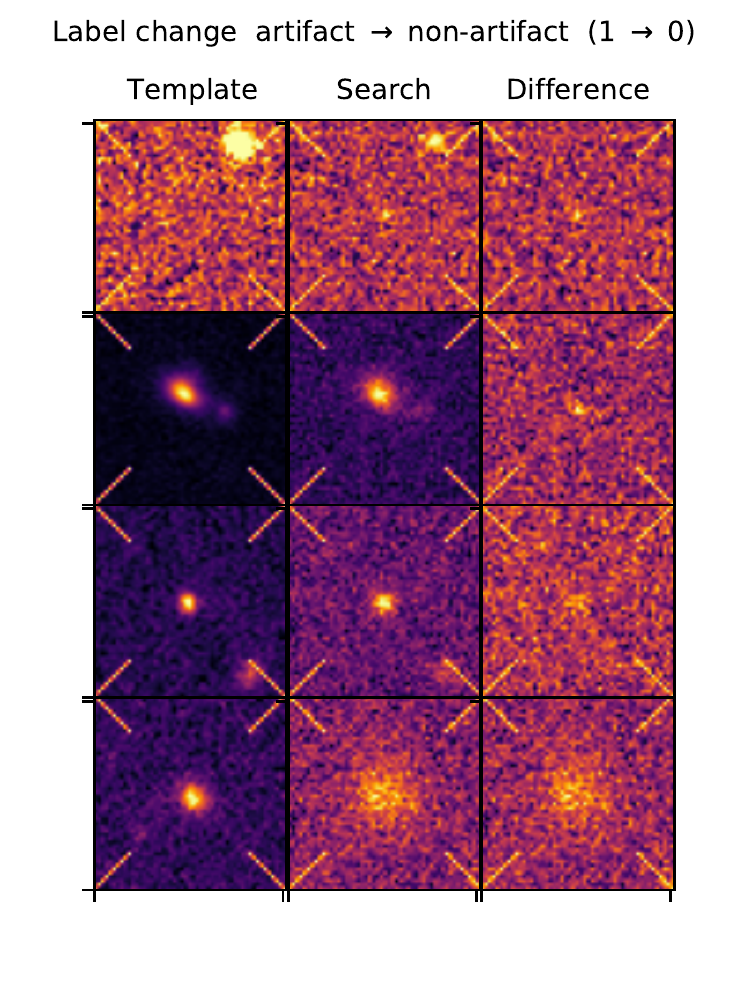}
\end{minipage}
\caption{ The left figure show four samples that were incorrectly labeled as non-artifacts, while  the right figure shows four samples that were incorrectly labeled as artifacts. In almost all cases of a True-Positive being mislabeled, it is due to the image stamp being placed on a saturated star or galaxy or on a bad part of the CCD. The mislabeling of the false negatives is well understood, and expected, as this sample in \citet{Goldstein_2015} was taken from all candidates and some true astrophysical transients crept into the data.}
\label{fig:relabel}
\end{figure}

The fact that different optimally trained CNN models strongly mis-categorize the same set of samples, points to the possibility of a case of mislabeling. To confirm this, we went back to the set of images in categories 1 and 6 for Model 1. Upon performing a visual inspection for a small subset of these samples, we found that about 90\% of these samples were indeed given the wrong label during the process of data preparation. As the artifact sample was randomly drawn from all detections, while the non-artifact came from injected transient candidates this is not too surprising. In fact, what we saw in our visual inspection is that some of the injected non-artifacts fell on bad parts of the CCD detector or were located near saturated stars while several of the artifacts were in fact heretofore unknown astrophysical transients.  This can be seen in Figure~\ref{fig:relabel}, where we show falsely classified non-artifacts in the left figure and falsely classified artifacts in the figure to the right.

\begin{figure}[ht!]
\plotone{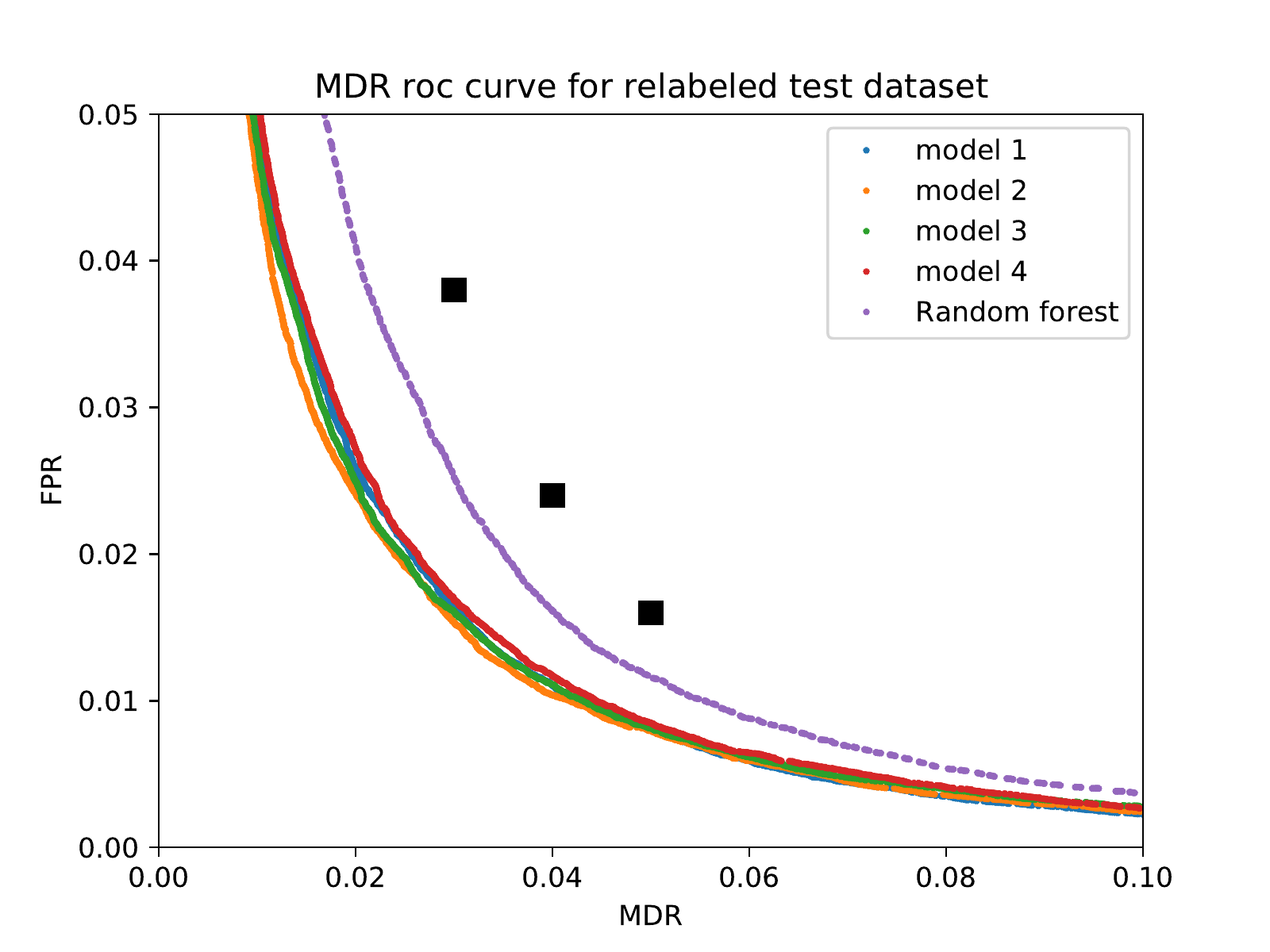}
\caption{The ROC curve of FPR vs MDR for the previously trained models using new labels for the test dataset. The black squares show the points obtained in \cite{Goldstein_2015} with random forest. The CNN models and random forest show significant improvement.}
\label{fig:roc_new_test}
\end{figure}

For the purposes of relabeling, we developed a GUI tool in python using the {\it Tkinter}~\citep{tkinter} library that enabled us to view blocks of images with their labels and allow an expert to quickly mark the images that require relabeling. 

As a first step, we performed a relabeling by inspecting roughly 750 samples that were classified into categories 1 and 6 by Model 1. Using the same trained models and their predictions, and just using the new labels, the resulting ROC curves are shown in Figure~\ref{fig:roc_new_test}. It is clear that the models are doing significantly better with the newer labels. In addition, the CNN models are now performing better than the random forest. Thus despite all models being trained with a dataset containing some mislabeled points, the CNNs are doing a better job at classification.

Having convinced ourselves of the efficacy of the relabeling process, we then obtained the predictions of Model 1 on the entire dataset.  Collecting the samples in categories 1 and 6, we then inspected this subset of 8093 samples. We found that 7402 (91\%) of these had to be relabelled. In all, 0.8\% of the samples were relabelled (7402 out of 898,963 samples).

\subsection{Results with new labels}\label{sec:resultswithnewlabels}

\begin{figure}[ht!]
\plotone{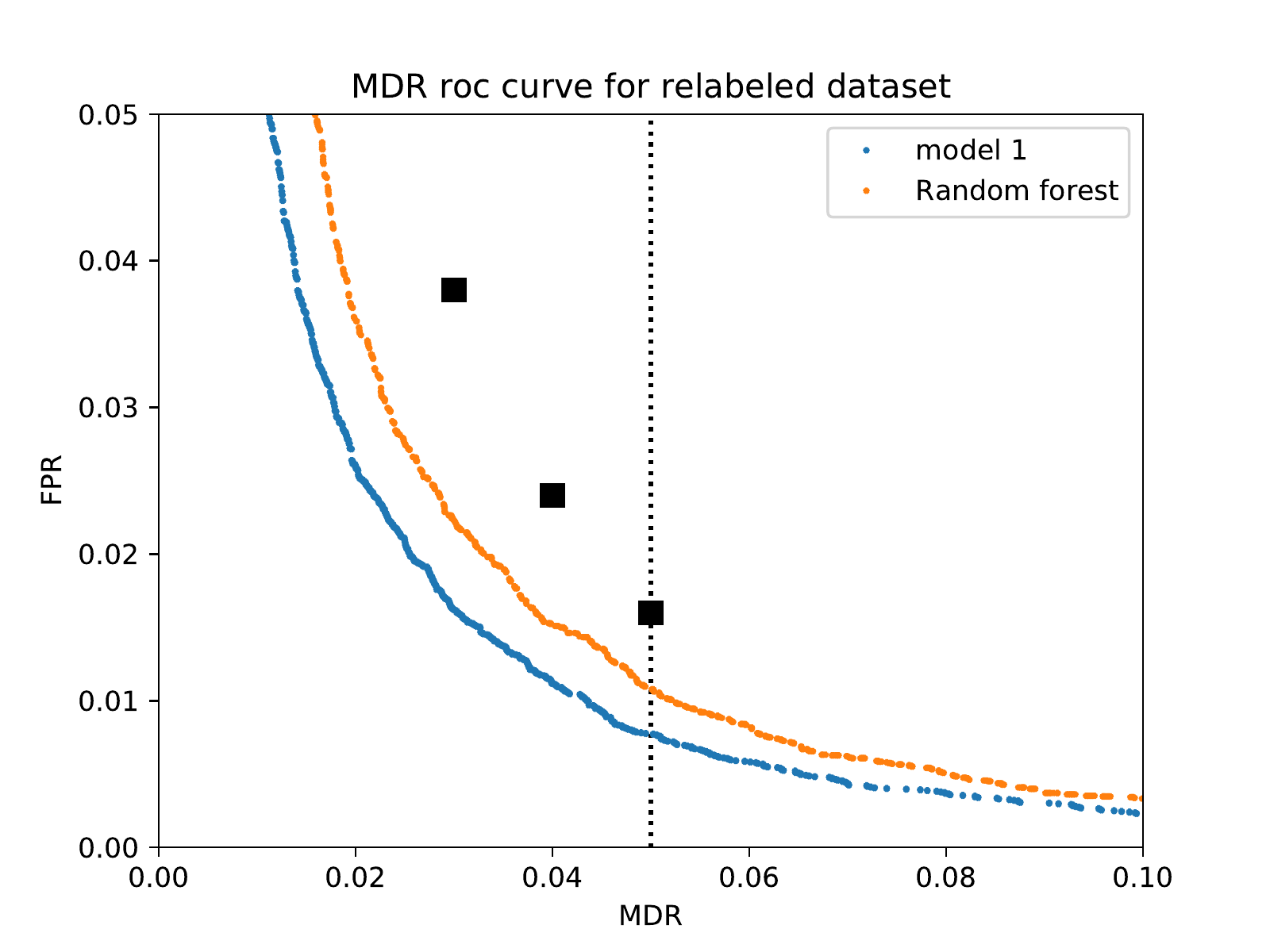}
\caption{The ROC curves of the best chosen CNN model: model 1 and random forest that were trained on the relabeled dataset. The black squares show the points obtained in \cite{Goldstein_2015} with the random forest. It is clear that Model 1 outperforms the random forest even on the relabeled dataset.  The dotted line represents an MDR value of 0.05. For this MDR value, the correspond FPR values of model1 and random forest are 0.008 and 0.011 respectively. Thus the FPR value of the CNN model is only $ 73\% $ of the FPR value of the random forest. }

\label{fig:final_roc_mdr}
\end{figure}

\begin{figure}[ht!]
\plotone{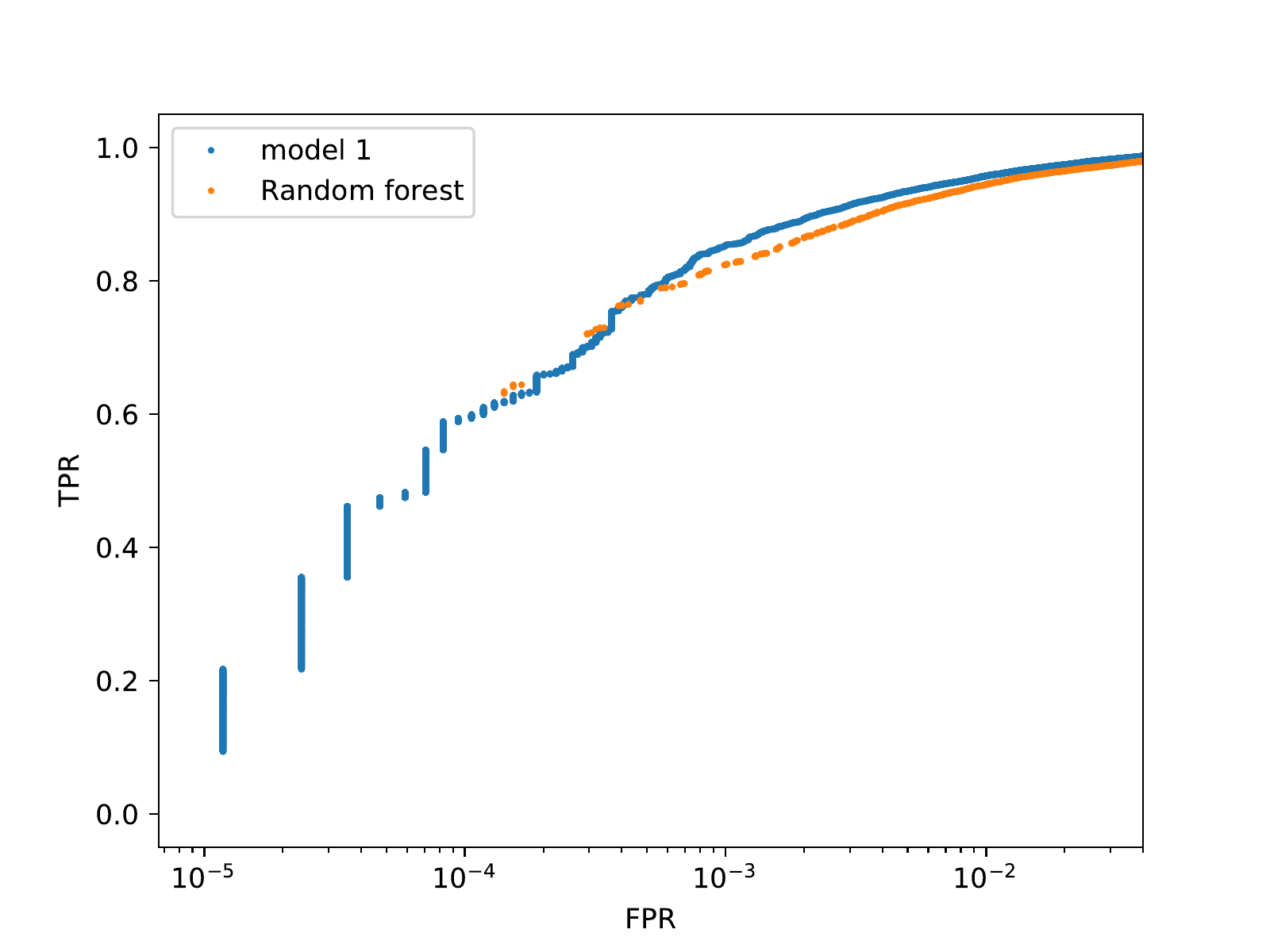}
\label{fig:final_roc_tpr}
\caption{The figure shows the ROC curves for Model 1 and random forest, plotting the true positive rate (TPR) with the false positive rate (FPR).}
\end{figure}

\begin{figure}[ht!]
\plotone{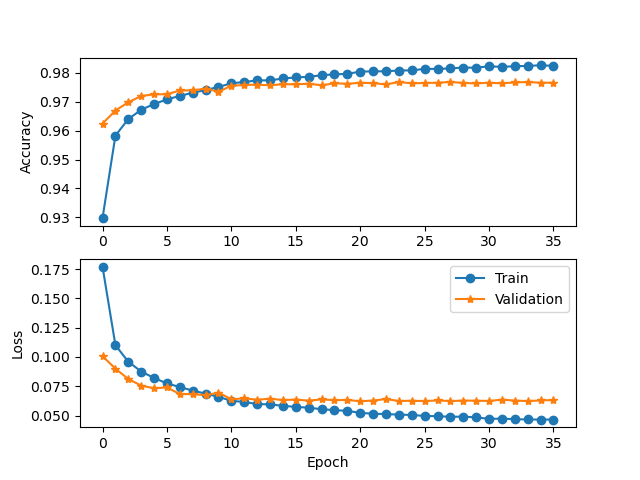}
\caption{The figure shows the learning curves for CNN Model 1. The top figure provides the training and validation accuracy values at the end of each epoch, while the bottom figure provides the corresponding loss values. The validation loss and accuracy stabilize after about 10 epochs.}
\label{fig:learning}
\end{figure}

\begin{table}[h!]
\begin{minipage}{.3\linewidth}
\centering
{ \bf Model 1}  \\
\vspace{0.1cm}
\begin{tabular}{|l|l|l||}
\hline
Layer & Output Shape & No. of Parameters  \\
\hline
\hline
Input       & $ 51 \times 51 \times 3 $ & 0 \\
Conv2D      & $ 51 \times 51 \times 80 $ & 2240 \\
BatchNorm   & $ 51 \times 51 \times 80 $ & 320 \\
MaxPooling  & $ 25 \times 25 \times 80 $ & 0 \\
Conv2D      & $ 25 \times 25 \times 80 $ & 57680 \\
BatchNorm   & $ 51 \times 25 \times 25 $ & 320 \\
MaxPooling  & $ 12 \times 12 \times 80 $ & 0 \\
Conv2D      & $ 12 \times 12 \times 80 $ & 57680 \\
BatchNorm   & $ 12 \times 12 \times 80 $ & 320 \\
MaxPooling  & $ 6 \times 6 \times 80 $ & 0 \\
Flatten  & $ 2880 $ & 0 \\
Dropout  & $ 2880 $ & 0 \\
Dense  & $ 51 $ & 146931 \\
BatchNorm  & $ 51 $ & 204 \\
Dense  & $ 1 $ & 52 \\
\hline
Total trainable & & \\ parameters & & {\bf 265,165} \\
\hline
\end{tabular}
\vspace{0.1cm}
\\ { \bf Model 2} \\
\vspace{0.1cm}
\begin{tabular}{|l|l|l||}
\hline
Layer & Output Shape & No. of Parameters  \\
\hline
Input       & $ 51 \times 51 \times 3 $ & 0 \\
Conv2D      & $ 51 \times 51 \times 80 $ & 3920 \\
BatchNorm   & $ 51 \times 51 \times 80 $ & 320 \\
Conv2D      & $ 51 \times 51 \times 80 $ & 102480 \\
BatchNorm   & $ 51 \times 51 \times 80 $ & 320 \\
MaxPooling  & $ 17 \times 17 \times 80 $ & 0 \\
Conv2D      & $ 17 \times 17 \times 80 $ & 102480 \\
BatchNorm   & $ 17 \times 17 \times 80 $ & 320 \\
Conv2D      & $ 17 \times 17 \times 80 $ & 102480 \\
BatchNorm   & $ 17 \times 17 \times 80 $ & 320 \\
MaxPooling  & $ 5 \times 5 \times 80 $ & 0 \\
Flatten  & $ 2000 $ & 0 \\
Dropout  & $ 2000 $ & 0 \\
Dense  & $ 51 $ & 102051 \\
BatchNorm  & $ 51 $ & 204 \\
Dense  & $ 1 $ & 52 \\
\hline
Total trainable & & \\parameters & & {\bf 414,205} \\
\hline
\end{tabular}
\end{minipage}
\begin{minipage}{.8\linewidth}
\centering
{\bf Model 3}  \\
\vspace{0.1cm}
\begin{tabular}{|l|l|l||}
\hline
Layer & Output Shape & No. of Parameters  \\
\hline
\hline
Input       & $ 51 \times 51 \times 3 $ & 0 \\
Conv2D      & $ 51 \times 51 \times 120 $ & 5880 \\
BatchNorm   & $ 51 \times 51 \times 120 $ & 480 \\
Conv2D      & $ 51 \times 51 \times 120 $ & 230520 \\
BatchNorm   & $ 51 \times 51 \times 120 $ & 480 \\
MaxPooling  & $ 17 \times 17 \times 120 $ & 0 \\
Conv2D      & $ 17 \times 17 \times 120 $ & 230520 \\
BatchNorm   & $ 17 \times 17 \times 120 $ & 480 \\
Conv2D      & $ 17 \times 17 \times 120 $ & 230520 \\
BatchNorm   & $ 17 \times 17 \times 120 $ & 480 \\
MaxPooling  & $ 5 \times 5 \times 120 $ & 0 \\
Flatten  & $ 3000 $ & 0 \\
Dropout  & $ 3000 $ & 0 \\
Dense  & $ 51 $ & 153051 \\
BatchNorm  & $ 51 $ & 204 \\
Dense  & $ 1 $ & 52 \\
\hline
Total trainable & & \\ parameters & & {\bf 851,605} \\
\hline
\end{tabular}
\vspace{0.3cm}
\\ { \bf Model 4} \\
\vspace{0.1cm}
\begin{tabular}{|l|l|l||}
\hline
Layer & Output Shape & No. of Parameters  \\
\hline
\hline
Input       & $ 51 \times 51 \times 3 $ & 0 \\
Conv2D      & $ 26 \times 26 \times 40 $ & 4360 \\
BatchNorm   & $ 26 \times 26 \times 40 $ & 160 \\
Dropout     & $ 26 \times 26 \times 40 $ & 0 \\
Conv2D      & $ 13 \times 13 \times 60 $ & 86460 \\
BatchNorm   & $ 13 \times 13 \times 60 $ & 240 \\
Dropout     & $ 13 \times 13 \times 60 $ & 0 \\
Conv2D      & $ 13 \times 13 \times 80 $ & 172880 \\
BatchNorm   & $ 13 \times 13 \times 80 $ & 320 \\
Dropout     & $ 13 \times 13 \times 80 $ & 0 \\
Flatten  & $ 13520 $ & 0 \\
Dropout  & $ 13520 $ & 0 \\
Dense  & $ 51 $ & 689571 \\
BatchNorm  & $ 51 $ & 204 \\
Dense  & $ 1 $ & 52 \\
\hline
Total trainable & & \\ parameters & & {\bf 953,785} \\
\hline
\end{tabular}
\end{minipage}
\caption{ Structures of the 4 best CNN models. As mentioned in~\ref{dataset}, each CNN reads a batch of input images of dimensions $ 51 \times 51 \times 3 $, with the 3 corresponding to the three types of images {\it Template, Search and Difference}. As the different layers of the CNN are applied to each  image, the dimensions of the image array change. 
The above table lists these details, with the first column describing the layer and the second column denoting the dimensions of the intermediate image array. The third column gives the number of parameters in each layer. The terms Layers Conv2D, Batch Norm, MaxPooling, Flatten, Dropout and Dense represent the standard CNN operations   convolution, batch normalization, maxpooling, flattening, dropout and dense respectively. More information about these can be found in the keras layers API \href{https://keras.io/api/layers/\#convolution-layers}{documentation}. 
}

\label{tab:model_structure}
\end{table}

\begin{figure}[ht!]
\plotone{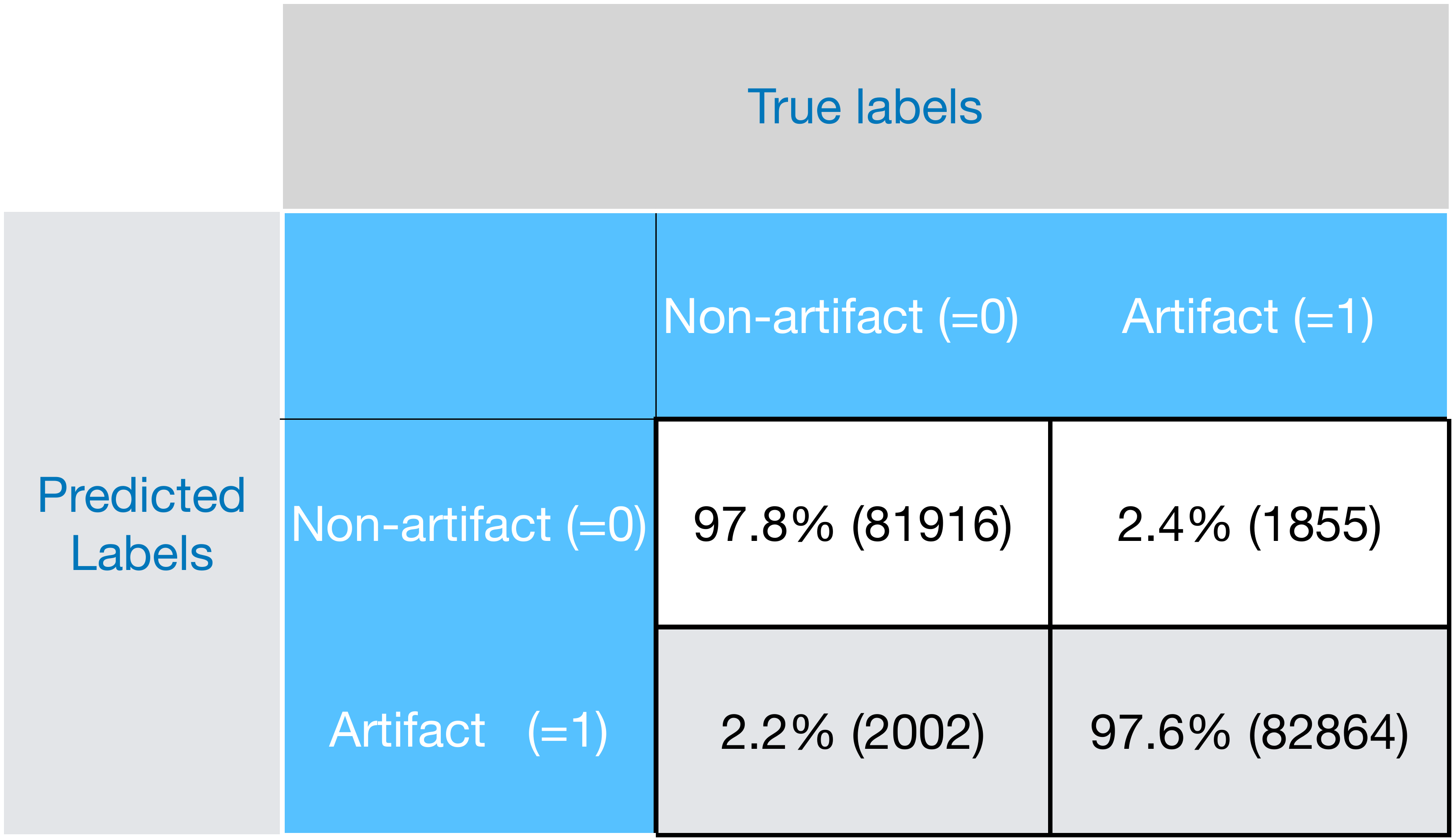}
\caption{The confusion matrix, with both normalized percentages and total number of triplets for used in training (in parens), for Model 1 based on a threshold cut of 0.5. }
\label{fig:confmat}
\end{figure}


\begin{table}[h!]
\centering
\begin{tabular}{|l||c|c|c|}
\hline
Category & {\bf 1}: Train and test with old labels &  {\bf 2}: train with old labels,test with new labels & {\bf 3}: train-test with new labels  \\
\hline
1 &  0.35 \% & 0.2 \% & 0.36 \% \\ 
2 &  0.75 \% & 0.82 \% & 0.83 \% \\ 
3 & 48.4 \% & 49 \% & 49.1 \% \\ 
4 &  48.4 \% & 48.8 \% & 48.5 \% \\ 
5 &  1.2 \% & 1.02 \% & 0.77 \% \\ 
6 &  0.84 \% & 0.23 \% & 0.33 \% \\ 
\hline
\end{tabular}
\caption{Comparing the points in various categories for the three cases: \\ 1: Train and test with old labels, 2: train with old labels, but test with new labels, 3: train and test with new labels. Comparing columns 1,2 and 3, it can be seen that column 2 has few points in categories 1 and 6, since the test dataset was biased by Model 1. Column 3 has slightly higher values in these categories compared to column 2, since the models were re-trained on a bigger dataset. Overall, column 3 has fewer points than column 1 in category 6, implying that the relabeling procedure had the intended effect. }
\label{tab:model1_comp}
\end{table}

We trained all four CNN models and the random forest on the relabeled dataset, after splitting the data into training (70\%), validation (10\%), and test samples (20\%). The resulting ROC curves are shown in Figure~\ref{fig:final_roc_mdr}. It is clear that the random forest is performing better with the new, relabeled dataset, and the four CNN are comparable in performance, but better than the random forest. Figure~\ref{fig:final_roc_tpr} shows the ROC comparing the true positive rate with the false positive rate as defined in Eqn~\ref{eq:tpr}. Based on the two ROC curves, we choose Model 1 as our best model, although the classification of the four CNN models are fairly similar. 
 To quantify the improvement in performance, we compare the FPR values for a fixed MDR value of 0.05 as shown in the Figure~\ref{fig:final_roc_mdr}. The corresponding FPR values for the CNN model 1 and random forest at 0.008 and 0.011 respectively. Thus the CNN model 1 lowers the FPR value by 27\%.

We present the learning curve of Model 1 in Figure~\ref{fig:learning} and its detailed structure in Table~\ref{tab:model_structure}. The confusion matrix is presented in Figure~\ref{fig:confmat}.

We would like to reiterate that the relabeling procedure has been conducted by visual inspection, with the machine learning method being used to only shortlist the suspected mislabeled images.
It might be argued that our relabeling process might be biasing the performance of Model 1, since we chose the points to relabel, based on the predictions of Model 1. However, in Section ~\ref{subsec:orig}, we trained with only 50\% of the samples, keeping the rest of the data in reserve. For this final analysis, we built the dataset with a different random seed and used 70\% of the data for training, thus mitigating the bias.
Table \ref{tab:model1_comp} shows the number of points in various categories for Model 1 for the three cases: train-test with old labels, train with old label, test with new labels, train and test with new labels. It is clear that values in column 2 are lowest, due to the bias. But the values in column 3 are intermediate, indicating that the bias has been mitigated. Hence, we are confident that our final CNN models are indeed better at classification than the random forest.

\subsection{Results in an ongoing survey}

We are currently using the CNN to provide a real/bogus score for an ongoing survey, the DECam Deep Drilling Fieds (DDF) program (Graham et al., 2022, in preparation). This is being run at the Blanco 4m telescope at Cerro Tololo-Inter-American Observatory as part of the DECam Alliance for Transients (DECAT), a consortium of time-domain DECam programs.  We have been using real/bogus scores produced by the CNN to decide which detections are sent out in alerts.  Throughout 2021, we used the original CNN trained for this paper.  Starting in 2022, and ultimately for the analysis of the entire data from the survey, we will be using a retrained CNN as described below for extragalactic ($|b|>20^\circ$) events. Training of a CNN for galactic events is in progress.  

Because this is real, incoming data, rather than a simulation, we don't have the absolute truth as to what's a genuine astronomical detection on the difference image, and what's a subtraction artifact or other "bogus" event.  In order to produce an evaluation data set, several observers manually tagged events from this survey as "real" or "bogus".  Participating observers were all trained on examples of good and bad events.  The observers included two with decades of experience in vetting supernova candidates in searches like this, and three undergraduate student assistants. Each observer was given events randomly chosen from the few million events that had been found by the data pipeline.  Once several observers had rated enough events, they were given events randomly chosen from those that had already been rated by others.  In this way, we were able to build up a set of $\simeq25,000$ events that had been tagged by three or more observers.

It is important to emphasize that what we are trying to do here is different from what is described in the rest of this paper.  The development of the CNN described in most of this paper aimed to correctly identify simulated transient candidates.  Here, we are using the same CNN architecture in an attempt to reproduce in bulk, the messier process of human scanning of transient candidates from real data.  The training and validation sets cannot be as clean in this case.  Of the $25,000$ events tagged by three or more observers, we selected those where the number of observers in the majority was at least two greater than the number in the minority.  (Effectively, this means that for those only rated by three observers, the tags would have been unanimous.)  Of this subset, 
about $1,700$ were tagged by the majority as good and $19,000$ as bad.  There was a unanimous agreement on 75\% of the events that the majority deemed to be good; there was a consensus on 95\% of the events deemed to be bad by the majority.  The choice to use the majority tags rather than the consensus tags represents a greater emphasis on reducing missed detections as opposed to reducing false positives.

\begin{figure}[ht!]
\plotone{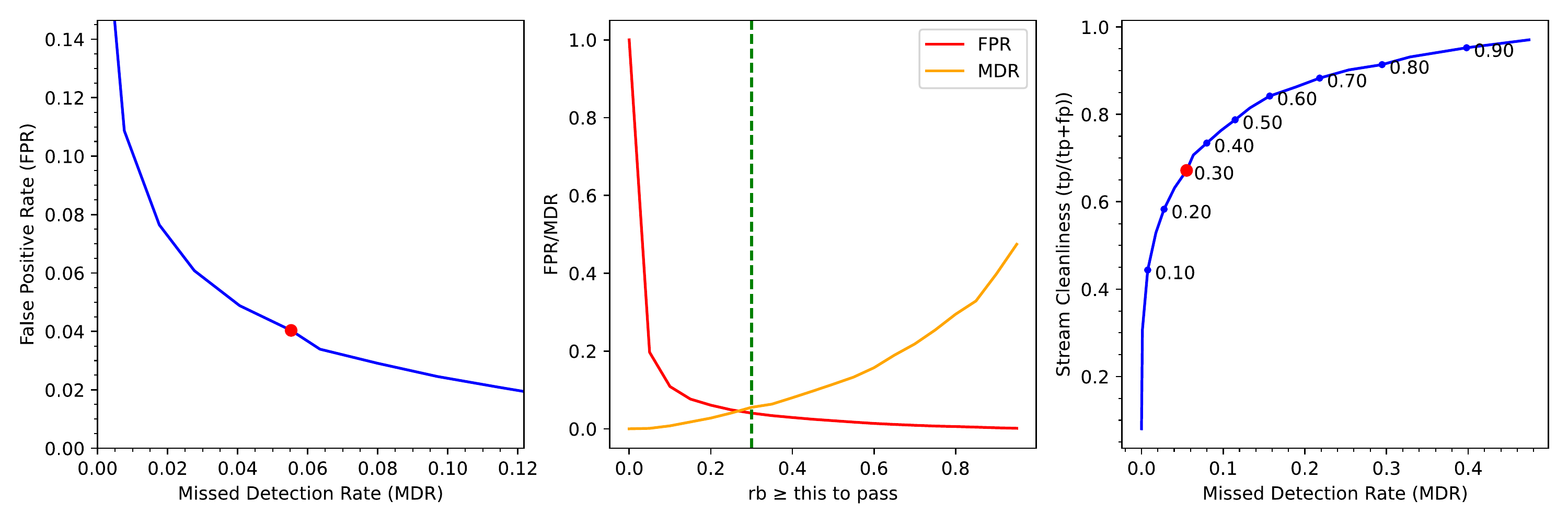}
\caption{Results for the CNN model trained against manual vetting of the ongoing DECAT/DDF survey.  Left: ROC curve.  Middle: false positive rate (FPR) and missed detection rate (MDR) as a function of real/bogus score cutoff.  Right: stream cleanliness vs. MDR.  Stream cleanliness is the fraction of passed objects that are real objects.  This is different from 1-FPR because in the real dataset, there are a factor of ~7 more bogus events than real events.  In the left and right plots, the red dot indicates the chosen real/bogus threshold of 0.3 that will be used in determining if an alert should be sent out for the detection.  The dashed green vertical line in the middle plot is the same cutoff.}
\label{fig:decatscorestats}
\end{figure}

\begin{figure}[ht!]
\plotone{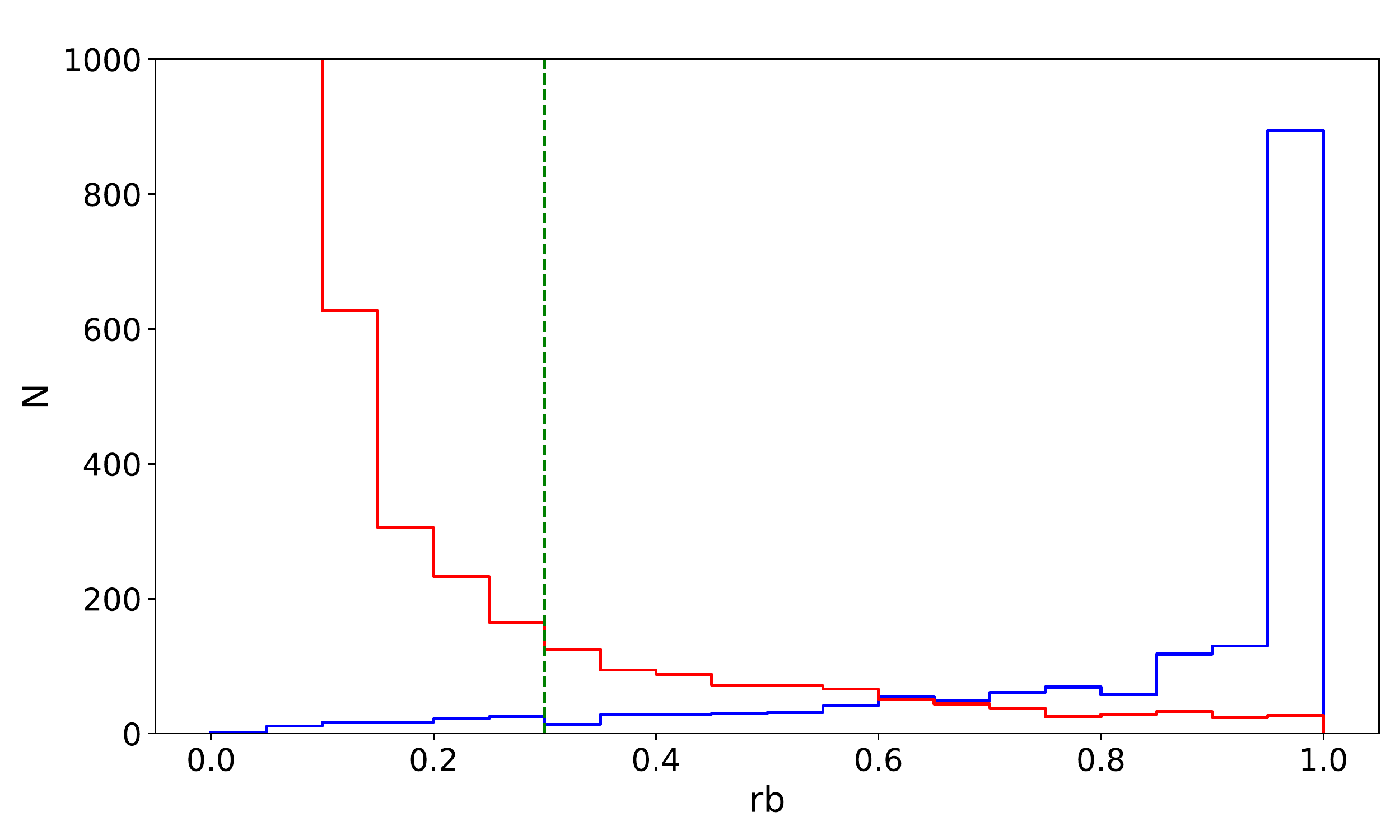}
\caption{Histogram of real/bogus produced by the retrained CNN for the ongoing DECAT/DDF survey.  The blue histogram are events labelled as "real" by at least two out of three visual inspections, and the red histogram are events labelled as "bogus".  The vertical dashed green line is the real/bogus threshold of 0.3 that will be used for generating alerts.  Compared to Figure~\ref{fig:hist1}; as discussed in the text, we cannot expect the histogram to be as cleanly separated with this dataset as we can for the dataset used for the bulk of the paper.}
\label{fig:decat_histogram}
\end{figure}

The CNN trained on the simulated transient candidates (whose results are in Section~\ref{sec:resultswithnewlabels}) did not perform particularly well in reproducing human scanning of the live data set.  In particular, there was a high missed detection rate (for any reasonable r/b cutoff) of $\sim0.5$ for candidates that were unanimously agreed to be good by three or more observers; this high MDR was present even when limiting to events with a high S/N ratio.  To allow the CNN to better model the visual scanning of this survey, we re-trained the model using the majority-tagged events described above as a training and validation set.  This retrained model performed much better than the original model on this new dataset, yielding a MDR of $\sim0.055$ and a FPR of $~\sim0.04$ (similar to the performance of the originally model).  The results for the retrained CNN are shown in Figures~\ref{fig:decatscorestats} and \ref{fig:decat_histogram}.  We cannot expect the ROC curve (left plot of Figure~\ref{fig:decatscorestats} here to be as good as the ones in Figure~\ref{fig:final_roc_mdr} because, as mentioned above, the training and validation set is not nearly as clean.

There is some evidence that the non-consensus events were at least sometimes more marginal cases, based on the r/b scores produced by the CNN retrained on the majority rankings.  The training of the CNN was just given a single real or bogus flag based on the majority of the human rankings; it had no knowledge of what was a consensus-good or consensus-bad event.  Despite this, the retrained CNN showed different statistics for consensus vs. non-consensus events.  For the majority-good events, the r/b scores of the consensus-good events were on average higher than the events on which there was disagreement (0.87 vs. 0.75).  For the majority-bad events, the r/b scores of the consensus-bad events were on average lower (0.05 vs 0.25).  Note that the fraction of events that had a consensus did not appreciably change when limiting to only high S/N events; those events that were marginal cases were not simply low-S/N cases, but represented cases where visual appearance of the residual might have had some suggestion of being an artifact, but was not clearly an artifact.

In conclusion, the model obtained using a CNN architecture optimized for the original dataset works fairly well with another, similar dataset, after a re-training of the weights. This points to our model architecture being fairly generic and hence more broadly applicable for datasets of this type. To improve upon the performance on this new dataset, we would have to perform another architecture search with a subset of this new data. We aim to address this in a future publication.

\section{Conclusions and Discussion} \label{sec:conclusions}

\subsection{Inference}
We have discussed automating the identification of transient detections obtained in astronomical imaging data using machine learning. Here we developed CNN models trained directly on the raw image data. The best CNN models match the performance of the previously used random forest method. In addition, using the CNNs predictions, we were able to identify that some of the images were mislabeled in the original data. After performing a relabelling of 0.9\% of the dataset, we re-trained the best CNN models. The resultant models outperform the original random forest method.
We also find that the CNNs are more robust to mislabeled samples in the training data.

Since we have only relabeled a small subset of the data, there is still the possibility that many other points are mislabelled. However, the significant increase in classification performance suggests that we have identified and relabelled most of the mislabeled images.

\subsection{Discussion}
There are two main benefits of using CNNs over random forest for image classification: classificiation efficiency and ease of use.

Currently in astronomy, random forest methods are the most common method to auto-identify transients in image subtractions.  In~\cite{10.1093/mnras/stv292}, the authors compared the performance of random forests with neural networks and demonstrated that their random forest was most efficient. However, in our work, we have demonstrated that our CNN, obtained by performing a detailed architecture search, outperforms our random forest.
 For example, comparing the FPR value for a fixed MDR, we find that the CNN model lowers the FPR by 27\% compared to the random forest.
Since the datasets are different, one cannot perform a direct comparison of ROC curves. However, the fact that our CNN outperforms the our random forest on the same dataset clearly demonstrates the benefit of using CNNs.

Recently \cite{Acero} used CNN's for transient discovery that is the most comparable to our work as both used the datasets from \cite{Goldstein_2015} to train the CNN's. While the focus of their paper is on performing transient discovery without image subtraction, they do present a confusion matrix for a similar design as ours. Both their false-positives and true-negatives are a factor of $\sim$2 larger than ours.

\cite{duev_ZTF} has created CNN's for the classification of transients in the Zwicky Transient Facility (ZTF). It should be noted that ZTF is slightly different than our survey in that $\sim$30\% of the images they take are undersampled \citep{Bellm_2018}, thus a direct comparison to our work is not exactly correct. That said, their confusion matrix is very similar to ours in quality (their true-positive's are slightly more pure while their false-negatives are slightly worse). As the number of validation triplets was low in their study, the uncertainties on these numbers are on the order of $\sim$2\%. Both these studies confirm the utility and benefit of using CNNs for transient candidate detection and that they are superior to random forest methods. 

Another benefit of CNNs is their ease of implementation. One of the drawbacks of the original random forest method is the need to identify a set of important features to use. This process is fairly painstaking, and also involves performing many, often computationally expensive, operations on the raw images. The CNN method, on the other hand works directly on the raw image data. Although it requires an architecture search, different models with reasonably high complexity perform well in classification. Their computational cost is also quite reasonable as they run efficiently on GPUs.

Hence, we believe the CNN method is more suitable for implementing automation of image subtraction classification in astronomy. Such methods could, and should be explored in upcoming transient surveys such as the Rubin Observatory~\citep{2020RNAAS...4...41S} and the La Silla Schmidt Southern Survey (Nugent {et al.}, in prep.) among others.

\section*{Acknowledgements}
This research used resources of the National Energy Research Scientific Computing Center (NERSC), a U.S. Department of Energy Office of Science User Facility operated under Contract No. DE-AC02-05CH11231. V.A.’s work was supported by the Computational Center for Excellence, a Computational HEP program in the Department of Energy’s Science Office of High Energy Physics (Grant \#KA2401022). P.E.N. and R.A.K. acknowledge support from the DOE under grant DE-AC02-05CH11231, Analytical Modeling for Extreme-Scale Computing Environments.

\bibliographystyle{aasjournal}
\bibliography{main.bib}

\end{document}